# SPECTRAL SIGNATURES OF ULTRA-RAPIDLY VARYING OBJECTS

Ermanno F. Borra,
Centre d'Optique, Photonique et Laser, and Centre de Recherche en
Astrophysique du Québec
Département de Physique, Université Laval, Québec, Qc, Canada G1K 7P4
(email: borra@phy.ulaval.ca)

## ABSTRACT

The main purpose of this article is to alert spectroscopists, particularly those involved in surveys, to the fact that rapidly pulsating sources induce periodic structures in spectra. This would allow the detection of new classes of objects sending bursts of pulses separated by constant time intervals that are too short to be detected with conventional techniques. The outstanding advantage of the technique is that there is no need for specialized instruments or surveys. One only must incorporate signal-searching algorithms into existing data analyzing software and use it with standard spectroscopic surveys, including existing ones. It is a small effort with a potentially huge pay-off because finding rapidly pulsating objects would be of enormous interest. Even a lack of detection could be used to eliminate exotic theoretical models.

## 1. INTRODUCTION

Astronomy is an observationally-led subject where chance discoveries play an important role (Fabian 2009). Fabian (2009) discusses at length the role that serendipity has historically played in Astronomy and remarks that the time domain is the least explored one in astronomy and continues to be rich in discoveries. Present techniques used to observe variable objects measure intensity variations and cannot detect extremely rapid time variations (shorter than nanoseconds). It is therefore possible that there exist some classes of exotic astronomical objects undergoing ultra-rapid periodic pulsations that have not yet been discovered. This may include signals from extraterrestrial intelligence. Astronomical spectroscopic surveys have become increasingly frequent. Most spectroscopists are probably not aware of the fact that periodic time variations of the intensity signal originating from a pulsating source modulate its spectrum with periodic structures. The main purpose of this article is to alert spectroscopists, in particular those carrying out spectroscopic surveys, to look for periodic structures in spectra. Ultra-rapid pulsators may also be detected from their peculiar colors in multi-color surveys.

## 2. SPECTRAL MODULATIONS FROM PULSATING OBJECTS

The theoretical analysis that follows is supported by the experiments of Chin et al. (1992). Using a grating spectrometer, they measured spectral modulation in the visible wavelength region caused by pairs of 150 femtosecond pulses separated by times varying between $3\ 10^{-13}$ seconds and $5\ 10^{-11}$ seconds. The pairs of pulses were periodically emitted at intervals of $1.3\ 10^{-8}$ seconds.

Consider a pulsating source sending a burst having a time dependent electric field $E(t)$ made of $N$ pulses $V(t)$ separated by a time interval $\tau$. It can be modeled by the convolution of $V(t)$ with a comb function $\sum_{m=1}^{N}\delta(t-t_m)$

$$E(t) = V(t) \otimes \sum_{m=1}^{N}\delta(t-t_m)\ , \quad (1)$$

where $\delta(t-t_m)$ is the delta function and $t_m = (m-1)\tau$ with $m$ an integer number.

Let us follow a mathematical treatment, similar to the one used to model diffraction gratings that can be found in many optics textbooks (e.g. chapter 5 in Klein & Furtak 1986). The Fourier transform of $E(t)$ gives

$$H(\omega) = G(\omega)\sum_{m=1}^{N}e^{-i\omega(m\tau)}\ , \quad (2)$$

with $\omega=2\pi/\tau$.

Equation (2) reduces to

$$H(\omega) = G(\omega)e^{i(N-1)\omega\tau/2}\left[\sin(\omega N\tau/2)/(\sin(\omega\tau/2))\right]. \qquad (3)$$

Astronomical spectrographs detect the time average of $S(\omega)=H(\omega)H^*(\omega)$, where the asterisk indicates complex conjugation. Using Equation (3) we obtain

$$S(\omega) = S_1(\omega)\left[\sin(\omega N\tau/2)/(\sin(\omega\tau/2))\right]^2. \qquad (4)$$

The experiments of Chin et al. (1992) show the periodic modulation of $S(\omega)$ predicted for 2 pulses by Equation (4) with $N=2$ as

$$S(\omega) = S_1(\omega)\left[2\cos(\omega\tau/2)\right]^2. \qquad (5)$$

With increasing $N$ Equations (3) and (4) predict comb-like shapes having sharp peaks separated by $2\pi/\tau$. Figure 1 shows $S(\omega)$ for $N = 2, 5$ and $10$. Because the separations between peaks, as wells as their shapes, are independent of $\omega$, the structure of the signal and the separation between peaks will be the same at the very large $\omega$ values in optical spectra. The teeth of the comb become increasingly sharper as $N$ increases so that $S(\omega)$ will closely resemble a comb function made of $\delta$ functions for very large $N$. This is consistent with the fact that as $N$ tends to infinity, Equation 1 becomes the convolution of $V(t)$ with a *Shah* function $III(t-t_m)$. The Fourier transform of the convolution of two functions is equal to the product of the Fourier transforms of the functions. The Fourier transform of a *Shah* function is another *Shah* function with spacing inversely proportional to the spacing of the original *Shah* function (chapter 10 in Bracewell 1986). In practice, for very large $N$, the observed comb function will be made of teeth having the shape of the instrumental response function.

Equation (3), (4) and Equation (5) show that the periodic frequency spacing is inversely proportional to the time between pulses $\tau$. This would allow us to directly observe pulses with time separations shorter than $10^{-10}$ seconds with standard spectroscopic equipment in the visual-infrared regions of the spectrum. One could measure time separations order of magnitude smaller than the nanosecond time separation limits of current techniques (e.g. Howard et al. 2004).

This opens up the possibility of finding astronomical objects that pulsate with extremely short periods that would be undetectable with presently used techniques. It would also be possible to detect non-periodic sources that send bursts of periodic signals or periodic sources, having relatively long periods that emit bursts of short-period signals. An example of this would be a source that sends random bursts of pairs of pulses. It would be detected, provided the time between bursts is large, but $\tau$ small compared to the signal detection time of the equipment $\Delta t$. This is because, in practice, all instruments have an effective detection time $\Delta t$. For a grating spectrograph, it is given by the relation between the beam elongation in the spectrograph and its resolution $\delta\lambda$ (Abramson 1993) as

$$\Delta t = (1/c)\lambda^2 / \delta\lambda. \qquad (6)$$

The separation $\tau$ between the pulses must be shorter than $\Delta t$ for a signal to be detected. Equation (6) indicates that $\Delta t$ and $\tau$ are very small for spectrographs. For example, for the Chin et al. (1992) experiment, where the resolution $\delta\lambda$ was 0.01 nm, Equation (6) gives $\Delta t = 10^{-10}$ seconds. The spectral modulation in Chin et al, (1992) begins to be barely detectable for $\tau = 3 \cdot 10^{-10}$ seconds but gives a strong signal for $\tau < 10^{-10}$ seconds, while the spectral modulation is visible down to $\tau = 4 \cdot 10^{-13}$ seconds, in agreement with Equation 6. The modulation could be seen at smaller time scales if the spectrum of a single pulse $S_1(\omega)$ is wider. Note that Chin et al. (1992) claim that their experiments show that the mathematical model that assumes infinitely long Fourier components is not simply a mathematical tool and that the infinitely long waves actually exist. Abramson (1993) makes the opposite statement. If Chin et al. (1992) are right, then the criterion expressed by Equation (6) does not apply and the effect of pulses separated by $\tau > \Delta t$ would be detected.

Because the period between spectral maxima is inversely proportional to the time between pulses $\tau$, very short periods will generate spectral features widely separated in frequency. Because spectra are usually limited in frequency by the detector, it is possible that only a single sharp emission feature may be found if $N \quad 2$. It may then be mistaken as the emission spectral line of a peculiar object (e.g. a high-redshift quasar). In the case where $\tau$ is small and $N=2$, the spectral limit of the detector may not include an entire cycle and only give a curved spectrum.

Widely separated spikes may also be detected in wide-band photometric surveys if a single spike is detected, thus generating peculiar colors. An object detected in a single filter would be particularly suspicious.

The basic theory on which this article is based is validated by the experiments of Chin et al. (1992). One may however argue that the pairs of pulses in Chin et al. (1992) are coherent, since the splitting was done with a Michelson interferometer, while the pulses in an astronomical source may not be coherent. One then may go on arguing that while coherent light beams interfere, it is "well-known" that incoherent ones do not.

Firstly, let us note that exotic objects capable of generating ultra-rapid pulses may generate pulses coherent in the classical sense. Secondly the "well-known" statement of the opening paragraph, sometimes contained in optics books, is wrong. Interference between independent optical sources has been experimentally demonstrated since the paper by Magyar & Mandel (1963) who detected interference effects by using integration times shorter than the coherence time. The reason why classical interference experiments do not detect interference is that they use integration times considerably longer than the coherence time of the light used: They therefore time-average over times significantly longer than the coherence time. Because phase randomly fluctuates over times longer than the coherence time, the long-term time-average of the interference pattern is not modulated. However, coherence can be artificially induced by periodic pulsation, allowing interference to be detected even after time-averaging. This has been demonstrated by the experiment of Alford & Gold (1958) that detects interference between two beams originating from two orthogonal directions of the same periodically pulsating thermal source (an electric spark). For all essential purposes, the interfering beams in Alford & Gold (1958) originate from independent light sources. Coherence is artificially induced by pulsation because pulsation ensures that the 2 beams generated at orthogonal directions, hence incoherent in the classical sense, are emitted with synchronized phases. In an ultra-rapidly pulsating source, the constancy of $\tau$ would ensure phase synchronization. Givens (1961) gives a theoretical analysis of the Alford & Gold (1958) experiment. He also derives the real part of Equation 3 for $N = 2$, which is equivalent to the square root of Equation 5, using the shift theorem of Fourier analysis

Scattering effects from the interstellar medium can be severe for very distant objects (Howard et al. 2004) but are negligible for nearby objects. For distant objects they will modify the profile of the intensity pulse by inducing exponential tails and decreasing the height of the peak, thus reducing the contrast of the spectral modulation. The effect will depend on position and distance, being worse in the direction of the galactic plane and minimum near the galactic poles.

One also must worry about the atmosphere of the earth. Kelly, Young & Andrews. (1998) studied the temporal broadening of ultra-short space-time Gaussian pulses with applications in laser uplink/downlink satellite communication. They find that pulses on the order of 10 - 20 femtoseconds can broaden by more than 100% whereas pulses greater than 500 femtoseconds have negligible broadening. They found no difference between uplink and downlink. They found that broadening was independent of frequency. Consequently, broadening is important for 10 femtosecond pulses but becomes negligible for 100 femtoseconds and beyond. Broadening will be less important for data taken at astronomical observatories, usually located on mountaintops where density and humidity are lower than at sea level. There will be no broadening effect for data obtained with space telescopes (e.g. Hubble).

## 3. COMPARISON WITH OTHER OPTICAL SEARCH TECHNIQUES

Any novel technique must answer two questions: What are its advantages over existing techniques and what are its disadvantages?

The physical parameters of many astronomical objects vary in time. Variable objects are commonly detected by observing how intensity varies as a function of time. While this works well for sources that have sufficiently slow time variations, it becomes inefficient and eventually inadequate as the time scale of variation decreases. Optical searches for rapid variations in astronomical objects have been proposed and carried out, principally to search for extraterrestrial intelligence (Howard et al. 2004, Stone et al. 2005, Hanna et al. 2009).They search for optical intensity pulses detected with fast electronics. In practice, they are limited to nanosecond timescales and bright objects (a few tens of thousand nearby stars at most). The next generation of instruments capable of high time resolutions is vey complex and would require very large telescopes (Barbieri et al. 2007). Furthermore, detectors and electronics that measure intensity variations have fundamental limitations for they cannot detect time variations below certain limits.

This article proposes detection in the spectrum and not in the intensity versus time domain. As far as time scales are concerned, the spectral technique picks up at timescales where the intensity techniques give up (nanoseconds) and goes down to picoseconds and beyond.

While the current intensity searches that use rapid detectors (e.g. photomultipliers) can only observe one object at a time and are limited to a few tens of thousands bright objects, the spectral modulation technique can go considerably fainter and observe billions of objects, since the search is done in spectra from standard surveys (e.g. SDSS or upcoming LSST). Spectroscopic surveys observe the large number of objects that fit inside two-dimensional detectors (e.g. CCDs). They therefore make a more efficient use of telescope time.

A small advantage also comes from the fact that surveys do not suffer from readout noise because of the long integration times. This is a problem with the nanosecond times needed with intensity techniques.

However, the outstanding advantage of the technique is its simplicity. There is no need for specialized instruments or surveys. All one has to do is to incorporate simple signal-finding algorithms that detect the type of spectral signature predicted by Equation 4 into existing software and use them with existing databases and future spectroscopic surveys.

The main disadvantage of the technique is that it cannot detect random signals. Another disadvantage comes from the limit set by Equation (6). If the time between pulses is longer than the limit set by Equation (6), the spectral modulation is not detected. Intensity techniques also have an advantage for the detection of pulses superposed on a bright object with a constant flux, like a star, since the contrast will be higher.

## 4. DISCUSSION

Although serendipity is, at this point in time, the only justification for a search, it must make a minimum of physical sense; while at the same time one must not be too narrow-minded. For example, let us not forget that astronomical phenomena with periods of fractions of seconds were not known before 1950 and the discovery of Pulsars was totally unexpected. A detailed discussion of physical phenomena is beyond the scope of this article. We shall therefore only carry out a brief consideration of the physical limits involved.

One may question the basic physics responsible for the pulses, the energy requirements as well as the energy density. Clearly the experiments of Chin et al. (1992) show that rapid pulses can be generated so that there is no violation of basic physical principles. However, a potential problem comes from the fact that short time scales imply small volumes and consequently large energy densities for distant objects so that thermal sources and well-understood physics may have problems with this. However, let us first note that the pulsating objects may be within the solar system itself thereby relieving the energy requirement. Let us also note that the basic physics of several serendipitously discovered astronomical sources, that also have energy density problems, is not well-understood either. In particular, the 2 nanosecond radio bursts observed in pulsars by Hankins et al. (2003) are difficult to explain with conventional physics. At 2 nanoseconds we are near the lower limit of the time scales detectable in spectra (10 nanoseconds). Hankins et al. (2003) speculate on the physics involved. The extreme nanopulses they detected are particularly puzzling. They are unresolved at 0.4 ns resolution and have an implied brightness temperature $2 \times 10^{41}$ K. Even assuming a Lorentz factor of $10^2$-$10^3$, the implied brightness temperature is $10^{35}$-$10^{37}$ K. They conclude that exotic non-thermal mechanisms are obviously responsible. Note that the less than 0.4ns width of the nanopulses is only a factor of 10 less than the 0.1 ns separation detectable in optical spectra.

Another example of strong and rapid high-energy emissions that are not understood come from soft-gamma-ray repeaters (SGRs) also called magnetars. They are galactic X-ray stars that emit numerous short-duration (about 0.1 s) bursts of hard X-rays during sporadic active periods. Hurley et al. (2005) detected a giant flare from SGR 1806-20. In the first 0.2 s, the flare released as much energy as the Sun radiates in a quarter of a million years.

Finally, let us note that that extreme Lorentz factors can generate narrow pulses originating in relatively low energy densities by decreasing, in the reference frame of observation, the time scales in the reference frame of emissions.

In conclusion, although there presently no Astrophysical object that is expected to pulsate below the nanosecond time scales, there are no compelling reasons to deny that such sources may exist.

## 4. CONCLUSION

Because a periodic variation of the intensity of a light source generates a periodic modulation in its spectrum, it opens up the possibility of finding new classes of objects undergoing extremely rapid pulsations. Figure 1 gives the spectrum of a pulsating source. The signal could also be superposed on a continuous spectrum, thereby decreasing the contrast. Extremely rapid pulsators sending pulses with time separations of the order of $10^{-10}$ to $10^{-13}$ seconds could be detected. This claim is supported by the work of Chin et al. (1992) who found spectral modulations originating from pairs of pulses separated by times intervals between $3 \cdot 10^{-13}$ seconds and $5 \cdot 10^{-11}$ seconds. Pulsations with separations less than nanoseconds would however be difficult to detect because the resolution of the spectroscopic equipment would be insufficient to resolve the spectroscopic signature. One could also detect periodic sources, having relatively long periods that emit bursts of short-period signals. Besides periodically varying objects, one also could detect objects that send random bursts with periodic structures (e.g. pairs of pulses). It would also be possible to find and observe pulsating sources having varying periods, provided the variations are sufficiently slow.

Fabian (2009) discusses at length the important role of serendipity in Astronomy. Indeed, Serendipity is the only justification for carrying out a search for ultra-rapid pulsators in spectra. At this point in time there is no indication whatsoever for the existence of any astronomical object that could give the type of signal that could be detected in spectra and it probably would not be worth the effort if specialized surveys or instruments were needed. However, all one has to do is to incorporate signal-finding algorithms into existing software and use it with existing databases and future spectroscopic or multicolor surveys. It is a very small effort worth doing because finding rapidly pulsating objects would be of enormous interest. This would particularly be the case if the pulses originated from Extraterrestrial Intelligence. It is a small effort with a potentially huge pay-off. Let us not forget that pulsars were not expected either in 1960

Finally, even if no ultra-rapid detector is found, the lack of detection may be used to eliminate exotic theories.

**ACKNOWLEDGEMENTS**

This research has been supported by the Natural Sciences and Engineering Research Council of Canada.

**REFERENCES**

Abramson, N.H. 1993, Applied Optics, 32, 5986
Alford, W.P., & Gold, A. 1958, Am. J. Phys., 26, 481
Barbieri, C. et al. 2007, J. of Modern Optics, 54, 191
Bracewell, R.N. 1986, The Fourier Transform and Its Applications, McGraw-Hill, New York
Chin, S.L., Francois, V., Watson, J.M., & Delisle, C. 1992, Applied Optics, 31, 3383.
Fabian, A, C. 2009, arXiv:0908.2784
Givens, M.P. 1961, J. Opt. Soc. Am., 51,1030
Howard et al. 2004, ApJ, 613, 1270
Hankins, T.N., Kern, J.S., Weatherall, J.C., & Ellek, J.A. 2003, Nature, 422, 141
Hanna, D.S. et al. 2009, arXiv:astro-ph/0904.2230v1
Hurley, K. et al. 2005, Nature, 434, 1098
Kelly, D.E. , Young. C.Y., & Andrews, L. 1998, Proc. SPIE , 3266, 231
Klein, M. V., & Furtak, T. E. 1986, Optics, John Wiley & Sons, New York
Magyar, G., & Mandel, L. 1963, Nature, 198, 256
Stone, R.P.S., et al. 2005, Astrobiology, 5, 604

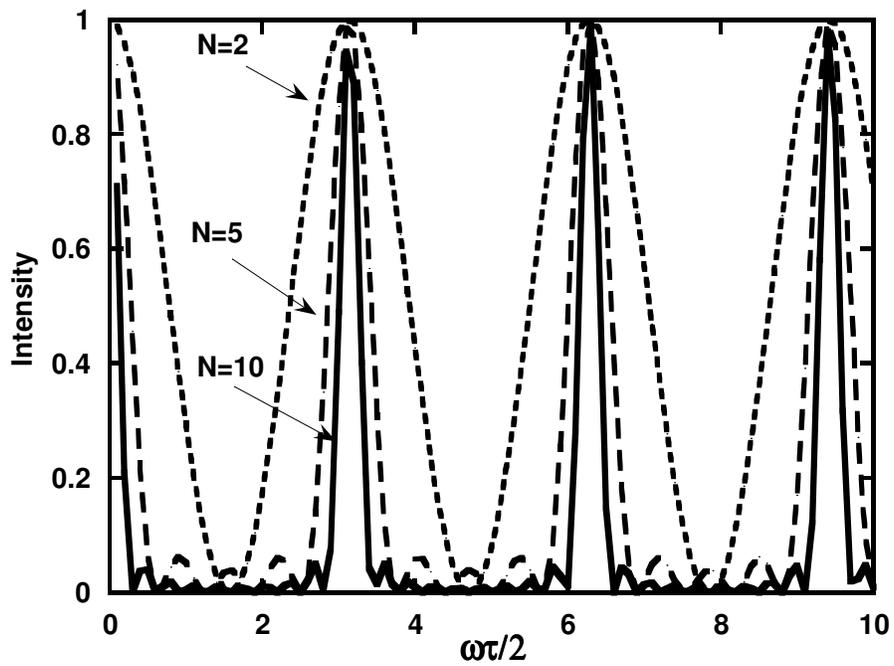

FIGURE 1: It shows $S(\omega)$, given by Equation (4), for $N = 2, 5$ and $10$. N gives the number of pulses that generated the spectrum and $\tau$ the time between pulses. This figure gives the spectrum of the pulsating source. The signal could also be superposed on a continuous spectrum, thereby decreasing the contrast.